\documentclass[twocolumn, prl]{revtex4} 

\usepackage{amsmath}
\usepackage[breaklinks=false]{hyperref}
\usepackage{color}
\usepackage{graphicx}
\begin{document}

\begin{@twocolumnfalse}
 \begin{center}
    \mbox{
{
\color{blue} \href{http://dx.doi.org/10.1364/OL.38.004935}{DOI: 10.1364/OL.38.004935}
}
    }
\end{center}
  \end{@twocolumnfalse}

\title{Focusing and imaging with increased numerical apertures\\ through multimode fibers with micro-fabricated optics}

\author{S. Bianchi,$^{1}$ V. P. Rajamanickam,$^2$ L. Ferrara, $^2$ E. Di Fabrizio,$^{3,4}$ C. Liberale,$^2$ and R. Di Leonardo$^{1,5}$}

\affiliation{
$^1$Dipartimento di Fisica, Universita' di Roma Sapienza, 00185, Rome, Italy\\
$^2$Nanostructures, Istituto Italiano di Tecnologia, Via Morego 30, 16163 Genova, Italy\\
$^3$KAUST (King Abdullah University of Science and Technology), PSE and BESE divisions, Jeddah, Saudi Arabia\\
$^4$BioNEM Lab, Dipartimento di Medicina Sperimentale e Clinica, Universita' Magna Graecia di Catanzaro, 88100 Catanzaro, Italy\\
$^5$National Research Council of Italy-IPCF UOS Roma, 00185, Rome, Italy
}

\begin{abstract}
The use of individual multimode optical fibers in endoscopy applications has the potential to provide highly miniaturized and non invasive probes for microscopy and optical-micro manipulation. A few different strategies have been proposed recently but they all suffer from an intrinsic low resolution related to the low numerical aperture of multimode fibers.
Here we show that two-photon polymerization allows for direct fabrication of micro-optics components on the fiber end resulting in an increase of the numerical aperture to a value that  is close to 1. Coupling light into the fiber through a spatial light modulator, we were able to optically scan a submicron spot (300 nm FWHM) over an extended region facing the opposite fiber end. Fluorescence imaging with improved resolution is also demonstrated.
\end{abstract}

\maketitle

Optical fibers can be used to deliver and collect light signals through non transparent media like biological tissues. In medical applications, optical fiber bundles are widely used in endoscopes, where  each single fiber carries the light intensity of a single pixel. A more compact endoscope for fluorescence microscopy applications can be built by mechanically scanning the tip of a single mode fiber, that provides the excitation spot, while the fluorescence signal is gathered by a multimode fiber. This last approach allows to fabricate endoscopes with probe diameters of few millimeters \cite{Wu2009endoscopy,Helmchen2011endo,Jung2003endo}. More recently, fully optical imaging techniques that employ a single multimode fiber (MMF) and require no moving parts are emerging. When using multimode fibers, however, any information at the input is encoded in the complex fields of their many propagating modes and then scrambled during propagation.  As a result, the output light distribution usually appears as a speckle pattern. In a first approach, a Spatial Light Modulator (SLM) has been used to control the phases of different modes so that they interfere constructively onto one or multiple diffraction limited spots \cite{fibre1}. These spots can be optically scanned and used for optical micromanipulation \cite{Cizmar2011shaping,fibre2} or for fluorescence microscopy \cite{fibre2,Papadopoulos2013scanning}. Other proposed strategies require no scanning of the excitation probe. In \cite{Choi2012fiber}, the direct measurement of the fiber transmission matrix allowed to numerically retrieve the sample's reflectivity from the analysis of the back--propagated field under random  speckle illumination.  Alternatively, fluorescence imaging can be achieved by recording the total fluorescence signals that result from a sequence of pre--recorded speckle patterns \cite{Mahalati2013speckle}. All of these applications, however, suffer from an intrinsic resolution limit that is imposed by the low numerical aperture of available multimode fibers with high mode capacities (typically NA 0.2-0.5).
 
\begin{figure*}[ht]
\centerline{\includegraphics[width=17cm]{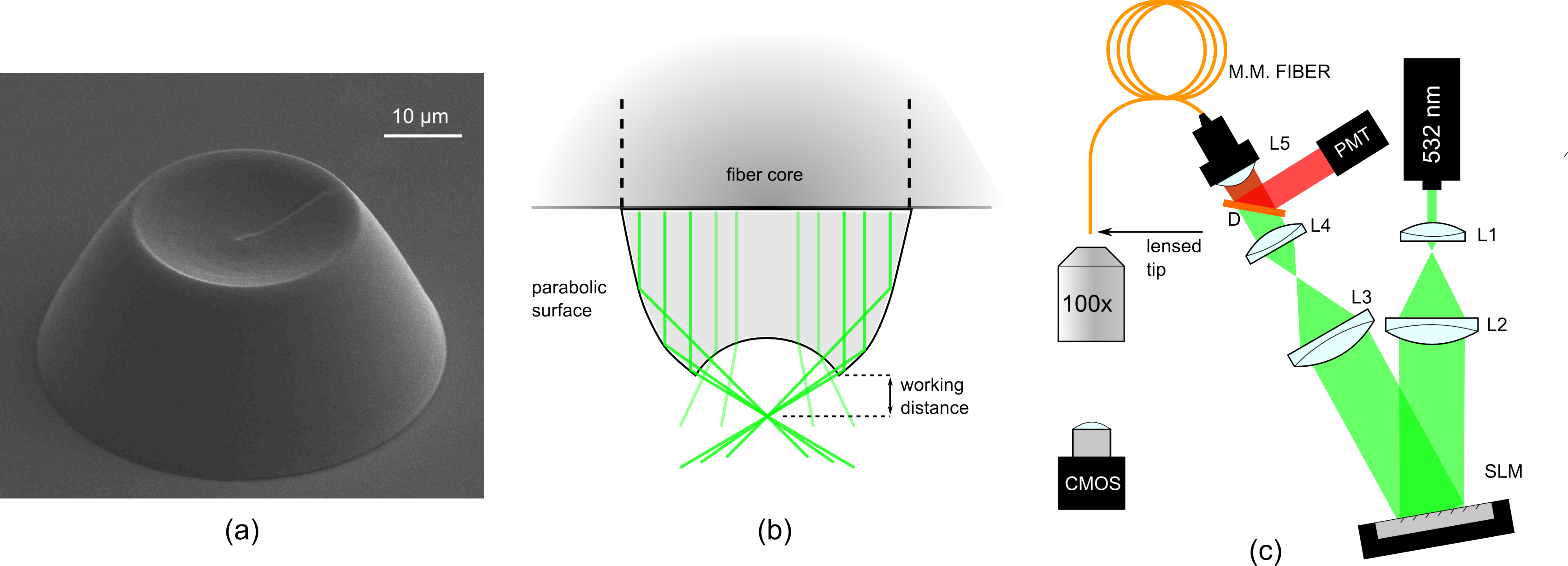}}
\caption{(a) A scanning electron microscope image of the parabolic micro reflector covering the core of a multimode fiber. (b) Ray optics description of the working principle of the micro-parabolic reflector. (c) Schematic view of experimental setup: L1-L4 planoconvex lenses (f=20 mm, 250 mm, 500 mm, 250 mm), L5 fiber collimator (f=11 mm, NA=0.25), D dichroic mirror, PMT photomultiplier}
\label{fig:setup}
\end{figure*}

\noindent A simple and effective way of introducing high spatial frequencies in the output field consists in having a thin layer of randomly packed small scatterers at the fiber output. Using this approach, sub-micron focusing was achieved in \cite{Papadopoulos2013increasing} and high resolution imaging was demonstrated in  \cite{Choi2013enhancement}. However, both studies reported a dramatic reduction in light transmission of more than 20 dB  due to strong scattering. 

Here we show that on-fiber miniaturized micro-optics can be used to extend the angular spectrum of outgoing light to substantially higher spatial frequencies. Using a SLM to shape light over one fiber end, we could optically scan a sub micron spot over the focal plane of a parabolic reflector that was fabricated directly on the opposite fiber end. Imaging with sub--micron resolution is also demonstrated. We used a parabolic reflector geometry that, besides achieving an extremely high numerical aperture, does not rely on refraction and therefore features a focusing power which is independent on the immersion medium. As an another important advantage in our approach, propagation through the micro-optical element only lowers the fiber transmission to 25\%. 

We used a Two-Photon Lithography (TPL) setup \cite{TPL_Carlo, TPL_Carlo2, TPL_Malinauskas} to fabricate a micro-parabolic reflector through point-by-point polymerization of a suitable photo-curable resin consisting of a thioxanthone photoinitiator and a diacrylate monomer. The fabricated micro-optics is shown in Figure \ref{fig:setup}(a), along with its description in the ray optics regime (Fig. \ref{fig:setup}(b)). Due to total internal reflection, light propagating inside the outer fiber core is reflected by the lateral parabolic surface. The central part of the structure has a spherical shape centered on the focal point of the parabolic surface. In this geometry reflected rays impinge at normal incidence on the spherical surface minimizing deflection due to refraction. The micro-optics has been fabricated to cover the core of a step-index MMF (Thorlabs AFS50/125Y, length=0.5 m, core size=50 $\mu$m, NA=0.22) with the following geometrical parameters: bottom and top diameters are respectively 50 $\mu$m and 30 $\mu$m, structure height is 26.5 $\mu$m, focal length and working distance are respectively 3.8 $\mu$m and 11 $\mu$m.  A small working distance is the price we have to pay when using a miniaturized optical system to achieve NA values that are close to 1.  By Fourier analysis of the generated output speckle pattern we estimated a cutoff (1/e$^2$) spatial frequency corresponding to an effective NA of 0.98. This increase in the NA comes with the cost of a reduced field of view which has approximately the same size of the top aperture of our structure.

A schematic view of our experimental setup is shown in Fig. \ref{fig:setup}(c). A laser beam (Coherent Verdi G, $\lambda$=532 nm) is expanded to fill the entire SLM (Holoeye LCR 2500) active area. The modulated beam is then compressed and focused on the input of the MMF. The optical train is aligned on the first diffraction order produced by adding a linear phase grating on the SLM. The zeroth order is focused outside of the fiber core and thus blocked. The micro-fabricated fiber tip is imaged by  100x objective (Nikon Plan Fluor, NA 1.3).

To test the improved focusing capabilities of our probe we followed the procedure described in \cite{fibre2}. The procedure consists in an initial calibration of the fiber field propagation \cite{Popoff2010transmission,Mosk2012review}, where a set of input modes are addressed in amplitude and phase while simultaneously measuring the intensities in the output modes. We choose as input and output coordinates two sets of field complex amplitudes $u_n$, $v_m$ evaluated on a square grid of sample points located respectively at the fiber input facet and at the focal plane of the parabolic reflector. From the linearity property of our system we have that:

\begin{equation}
\label{linearity}
v_m=\sum_n G_{mn}u_n
\end{equation}

\noindent The SLM is placed on the Fourier plane of the fiber input facet. When the input modes are addressed individually, by displaying linear phase gratings on the SLM,  we can obtain a direct measure of the propagator matrix amplitudes as $|G_{mn}|^2=|v_m|^2$. Using a random mask algorithm \cite{RM} we can simultaneously address two input modes with the same intensity and an arbitrary relative phase $\varphi$. One of these spots is kept fixed ($n=0$) and used as a reference while the second is scanned over the whole set of input modes. By successively setting the relative phases $\varphi$ to $0$,$\pi/2$,$\pi$ and $3\pi/2$  we can recover the arguments of $G_{mn}$ as:

\begin{equation}
\arg(G_{mn})=\arg(G_{m0})+\arctan\left[ \frac{I_m^{\pi/2}-I_m^{3\pi/2}}{I_m^{0}-I_m^{\pi}} \right]
\end{equation}

\noindent where $I_m^\varphi$ is the intensity on the output point $m$ when addressing input modes $0$ and $n$ with a relative phase $\varphi$. Having a speckle filed as a reference for finding the phase shifts,  results in considerable errors in $\arg(G_{mn})$ for those output points $m$ where the reference intensity is low. We can overcome this problem by using multiple reference fields and choosing the value for $\arg(G_{mn})$ that corresponds to the reference mode with the largest intensity in $m$ \cite{Cizmar2011shaping}. 
Once the propagation kernel is known, a possible choice for maximizing the field amplitude at the target output mode $m$  is given by $u_n=G_{mn}^*$ \cite{fibre2}. The required phase modulation is then obtained as the argument of the Fourier transform of $u_n$.

\begin{figure}[ht]
\centerline{\includegraphics[width=.48\textwidth]{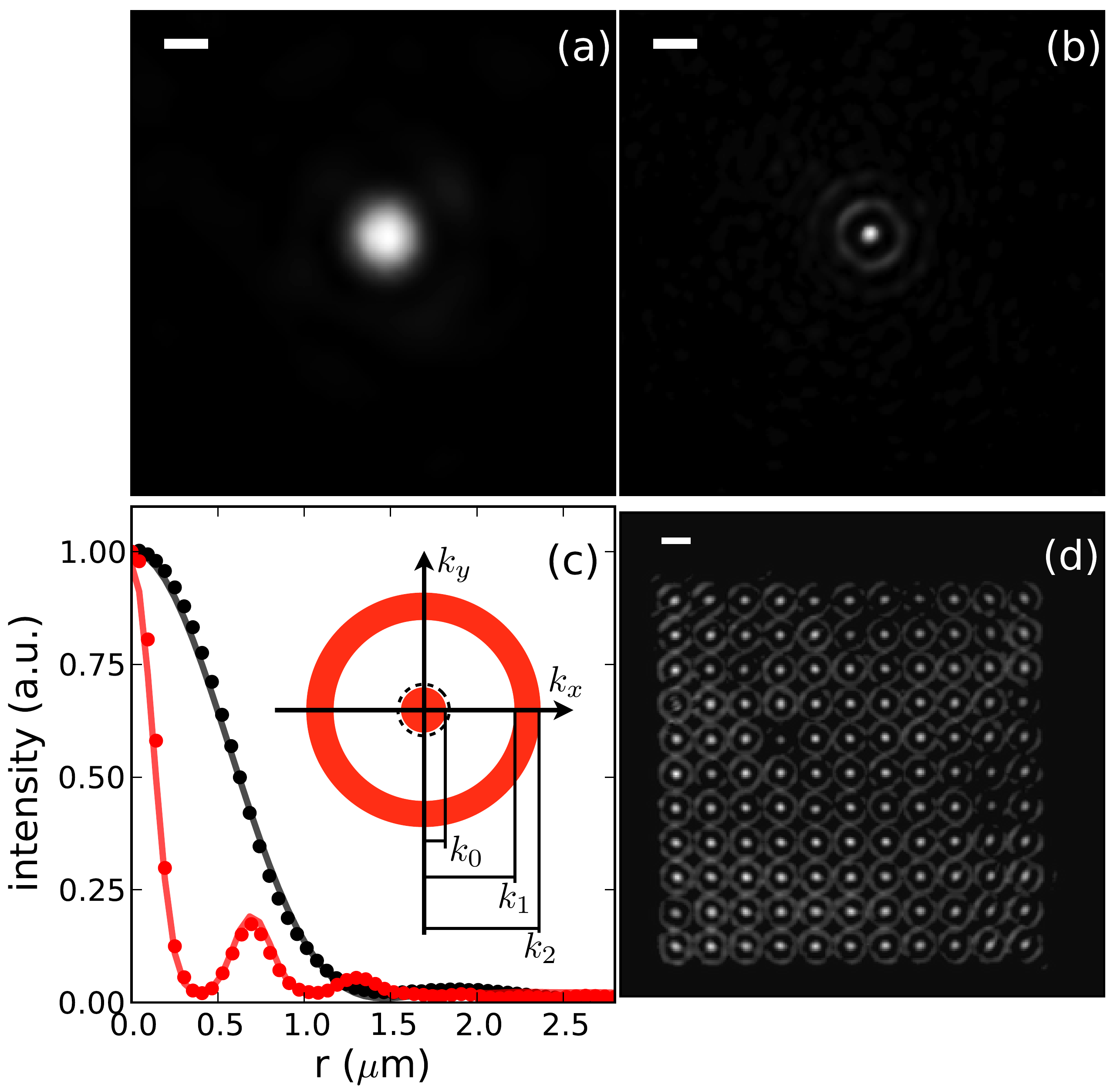}}
\caption{
Light refocusing at the output of a bare multimode fiber (a) and for the same fiber terminated with a parabolic reflector (b). (c) Radial profile of the focused spot in (a) (black circles) and (b) (red circles). The corresponding fits to an Airy pattern and Eq. \ref{spot} are reported respectively as a black and red solid lines. The inset shows a representation in reciprocal space of the spatial frequency content of the fitting spots. (d) Superposition of single spot images during an holographic scan over a square grid. All scale bars are 1 $\mu$m.}
\label{fig:spot}
\end{figure}

Fig. \ref{fig:spot}(a-b) shows the obtained focused spots coming out from a bare fiber (a) and from our structured fiber (b). The radial profiles of the two peaks, obtained from an azimuthal average, are plotted by solid circles in Fig. \ref{fig:spot}(c). For the bare fiber the spot's profile is fitted very well by an Airy pattern with NA=0.22 plotted as a solid black line. On the other hand, the spot produced with the structured fiber (red circles) can be represented as the coherent superposition of two contributions that add up in phase at the focal point (see inset in Fig. \ref{fig:spot}c). 
The first component is given by unreflected light propagating through the central part of the structure and whose transverse spectral representation is given by a disk of radius $k_0$, that is an Airy pattern.
The second component comes from light that is reflected by the parabolic surface and has an annular spectral representation with transverse wave vectors in the range $[k_1, k_2]$.  The intensity of the resulting spot is given by: 
\begin{equation}
\left|\frac{J_1(k_0 r)}{\sqrt{\pi}r}+w\frac{k_2 J_1(k_2 r)-k_1 J_1(k_1 r)}{r \sqrt{\pi(k_2^2-k_1^2)}}\right|^2
\label{spot}
\end{equation}
\noindent where $w$ is a weight factor. By a fitting procedure we obtain $k_0=2.13$ $\mu$m$^{-1}$, $k_1=8.56$ $\mu$m$^{-1}$, $k_2=11.1$ $\mu$m$^{-1}$ corresponding to numerical apertures of 0.18, 0.72, 0.93 respectively (see Fig. \ref{fig:spot}c). 
  
The spot can be holographically scanned across a 16$\mu$m$\times$16$\mu$m square area inscribed inside the reduced aperture of our structured fiber. The minimum scanning step is given by the spacing between the output M sample points which has been set to 220 nm in the present case. Figure \ref{fig:spot}(d) shows a superposition of sequential images where single spots are scanned over a square grid. Using three reference modes in the characterization of $G_{mn}$ the spots maxima have a standard deviation of 7\% over the entire scanning area.

\begin{figure}[ht]
\centerline{\includegraphics[width=.48\textwidth]{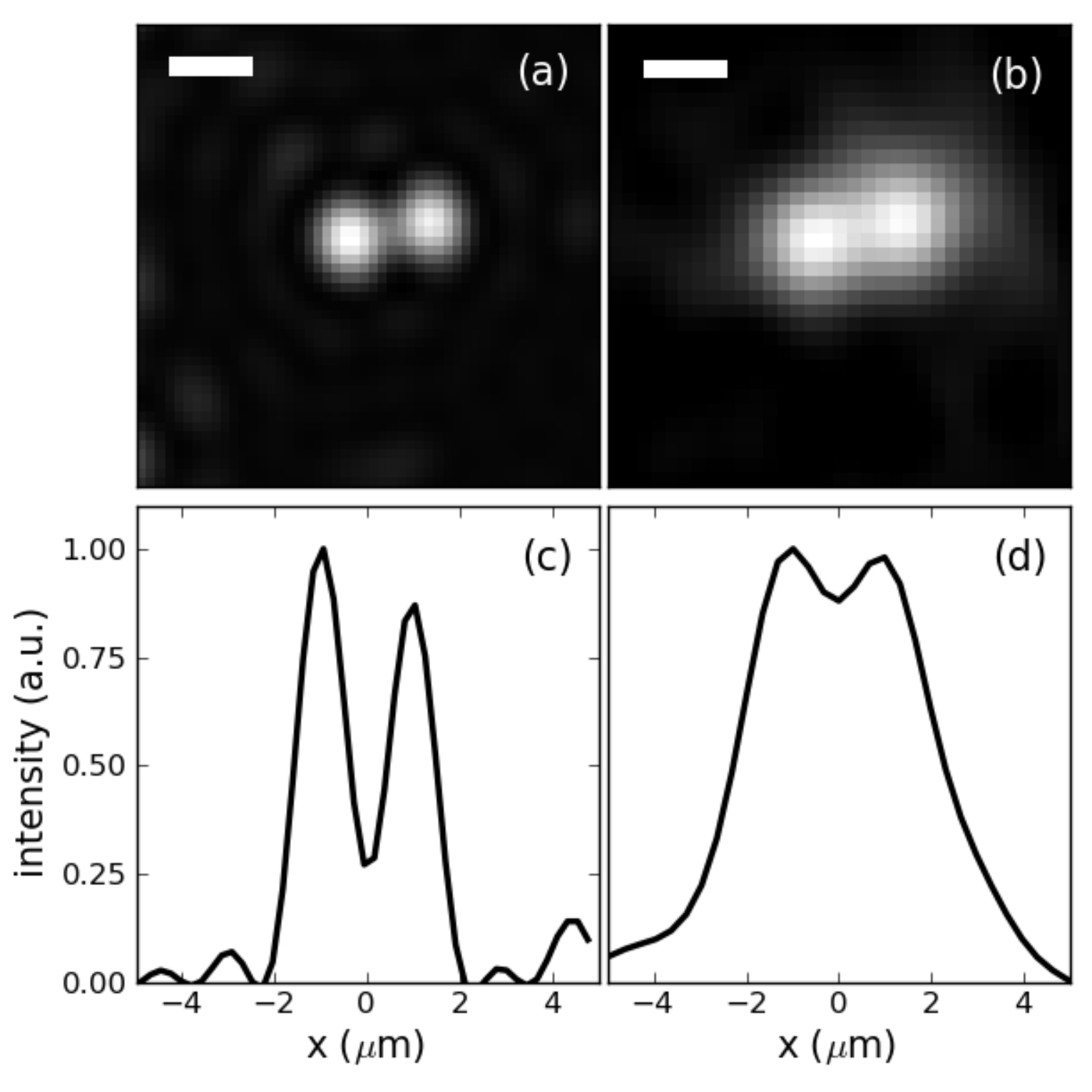}}
\caption{Fluorescence images of 2 $\mu$m polystyrene beads obtained with a structured fiber (a) and a bare fiber (NA=0.27) (b). The corresponding intensity profiles are also shown (c-d). All scale bars are 2 $\mu$m.}
\label{fig:fluo}
\end{figure}

The increased NA of our structured fiber can be also exploited  to achieve a higher resolution in fluorescence imaging.  The ability of forming diffraction limited spots requires a detailed knowledge of fiber propagation characteristics, and is crucial in applications like optical trapping. On the other hand, fluorescence imaging can  be achieved in a more simple and easy to implement approach that requires no need of wavefront modulation or interferometric phase retrievals.  Following \cite{Mahalati2013speckle}  we use a digital camera to record the set of $N\simeq1500$  speckle patterns that are produced at the fiber output while scanning the input spot over an array of $N$ predetermined positions inside the input fiber core. 
The intensity distribution in each of those speckle pattern is stored into an $M=100\times100$ pixel image $I_{mn}$. 
Subsequently the fiber approaches a fluorescent sample and projects the same sequence of speckles while simultaneously recording the fluorescence signal that propagates back  through the fiber.  Once the procedure is completed we end up having $N$ fluorescence counts  $f_n$ that are proportional to the overlap between the illumination patterns $I_{mn}$ and the density map of fluorescent molecules $x_m$:

\begin{equation}
f_n=\sum_m I_{mn}x_m
\end{equation}

\noindent or in more compact matrix form $\bf f=I\cdot x$.
A good estimate for the unknown image $\bf x$ can be obtained
through the singular value decomposition  $\bf{I=U\cdot D\cdot  V}^T$:

\begin{equation}
\label{pseudo}
\bf{x}=\bf{V}\cdot\bf{D}^{-1}\cdot\bf{U}^T\cdot\bf{f}
\end{equation}

so that ${\bf x}$  minimizes the quantity $|\bf{f}-\bf{I\cdot x}|^2$ \cite{recipes}. When the inverse of the diagonal matrix $\bf{D}$ is ill conditioned, it is a common practice to set to zero those components in  $\bf{D}^{-1}$ whose value exceeds a given threshold. A better estimate of the real image $\bf x$ can be found by using a priori information whenever available.  In our case we know that $\bf x$ is positive and has a limited bandwidth of spatial frequencies. This a priori information can be used in an iterative procedure \cite{recipes} that alternates between steepest descents along the gradient of $|\bf{f}-\bf{I\cdot x}|^2$ and projections onto the subspace of solutions that are compatible with our a priori information:

\begin{equation}
\bf{x}^{k+1}=\mathcal{P} \left[ \bf{x}^k-(\bf{I}^T\cdot \bf{I}\cdot \bf{x}^k-\bf{I}^T\cdot \bf{f})\epsilon \right]
\end{equation}

\noindent where $\epsilon$ is a step parameter multiplying the gradient of $|\bf{f}-\bf{I\cdot x}|^2$ while $\mathcal{P}$ is a projection operator that sets to zero the negative elements of $\bf{x}$ and its high frequency components.  At the first iteration $\bf{x}^0$ can be found using equation \ref{pseudo}.
We tested our technique on a sample of 2 $\mu$m polystyrene fluorescent beads. 
For this particular sample, we found that an optimal choice for the frequency cutoff, providing a good balance between resolution and low noise, corresponds to an effective NA of 0.65. 
Figures \ref{fig:fluo})(a-b) show reconstructed sample images using respectively a structured fiber and a bare one. Intensity profiles in Fig. \ref{fig:fluo}(c-d) show that, in contrast to the bare fiber case, the two particles can be resolved much clearly when using the structured fiber.

In conclusion, we have demonstrated that a single multimode fiber, with an original NA of 0.22, can be terminated with a micro-fabricated parabolic reflector that allows the fully optical  scanning of a sub-micron spot having an angular spectrum that extends up to a NA of 0.93. The increased NA of our fiber is also used to collect fluorescence images with a much better resolution than what previously reported for bare fibers. Our results can be relevant for the use of multimode fibers in endoscopy, proving that the achievable resolution is not limited by the NA of the bare fiber but significantly higher values can be obtained employing on-fiber micro-optics with small transmission losses.

The research leading to these results has received funding from the European Research Council under the European Union's Seventh Framework Programme (FP7/2007-2013) / ERC grant agreement n¡307940.  We also acknowledge funding from MIUR-FIRB Project No. RBFR08WDBE.


\end{document}